# An unscented Kalman filter method for real time input-parameter-state estimation

Marios Impraimakis[a,*], Andrew W. Smyth[a]

[a] *Department of Civil Engineering and Engineering Mechanics, Columbia University, New York, NY 10027, USA*

---

**Abstract**

The input-parameter-state estimation capabilities of a novel unscented Kalman filter is examined herein on both linear and nonlinear systems. The unknown input is estimated in two stages within each time step. Firstly, the predicted dynamic states and the system parameters provide an estimation of the input. Secondly, the corrected with measurements states and parameters provide a final estimation. Importantly, it is demonstrated using the perturbation analysis that, a system with at least a zero or a non-zero known input can potentially be uniquely identified. This output-only methodology allows for a better understanding of the system compared to classical output-only parameter identification strategies, given that all the dynamic states, the parameters, and the input are estimated jointly and in real-time.

---

*corresponding author
*Email address:* m.impraimakis@columbia.edu (Marios Impraimakis)

## 1. Introduction

Output-only strategies in modern dynamical system monitoring, using response-only sensors to track the inputs, the parameters, and the dynamic states of the system, attract the attention for two main reasons. The first is that engineers try to extract as much information as possible from the available data. The practical reason is that data acquisition instrumentation (sensors, actuators, conditioners etc.) are cost-ineffective, space-consuming, prone to faults or damage [1, 2] and, in the case of a real-time monitoring, require an expert team to work continuously on them.

The second reason is that, in some circumstances, the input cannot be measured or the measurement of the input is more unreliable than the dynamic state measurement. For instance, there is not a reliable means of accurate measure the traffic and wind load on large structural systems.

For both reasons, developments of data fusion and system identification algorithms with maximum information extraction should be introduced from possible response-only sensor networks to monitor the dynamic systems, detect damage, and potentially provide a reliable damage prognosis.

Along these lines, many authors have proposed procedures to tackle the unknown input problem (i.e. [3–8]). The vast majority of those methods however, assume known system parameters, a fact that is in opposition to the damage detection philosophy.

Towards this need, Dertimanis et al. [1] proposed a successive Bayesian filtering framework to address the joint input-parameter-state estimation problem. The method successfully identified the linear systems using also limited output information. According to the authors though, a more physically meaningful process for adjusting the covariance matrices of the adopted fictitious equations needs to be explored in order to support, if not substitute, the L-curve regularization procedure.

Castiglione et al. [9] proposed a time-varying auto-regressive model for the unknown inputs, and develop a strategy to simultaneously estimate them along with the parameters. However, the authors stated that due to the low frequency drift in the estimated input, displacement responses could not be estimated satisfactorily.

Maes et al. [10] proposed a linerazation for the adopted system model around the current state, yielding an algorithm similar to the extended Kalman filter. Though the method was applied successfully to linear systems, the implementation of the Jacobian derivative may have undesirable consequences in a real large system or, more so, to highly nonlinear systems.

Lei et al. [11] proposed a recursive nonlinear least squares strategy for the unknown input, within the unscented Kalman filter (UKF) concept. Nonetheless, the nonlinear least squares concept can possibly lead to ineffective noise filtering, and this may be the reason why the introduced state and input estimation diverges more and more as the vibration continues.

Song [12] proposed several versions of a minimum variance unbiased UKF with direct feedthrough, and the methodology was successful. The author though examined only inputs that are incorporated into the process error.

Lastly, offline system identification techniques without any input information have been developed [13–16], but apart from the parameter estimation they are often unable to provide the dynamic state and the input estimation.

In this work, the capabilities of a novel unscented Kalman filter is ex-amined. The unknown input is estimated in two stages within each time step. Firstly, the predicted dynamic states and the system parameters provide an estimation of the input. Secondly, the corrected with measurements (updated) dynamic states and parameters provide a final input estimation.

The proposed method can be used successfully for linear and nonlinear systems when a zero or a non-zero known input exists. Displacement, velocity, and acceleration signals are recommended for a full system estimation, while a data fusion methodology may relax this assumption. However this is not mandatory as it will be demonstrated using a sensitivity analysis.

Here, Jacobian derivatives, least square introductions, nonphysical propagation processes, or state estimation sacrifices will not be needed.

The work is organized as follows: in section 2 the well known and widely accepted unscented Kalman filter is mathematically presented, in the case of joint dynamic state and parameter estimation. The proposed methodology is developed in section 3. For both the standard and the proposed UKF, a Table with pseudo-code is provided. In section 4, the identifiability characteristics of the discussed problem are examined, while in section 5 numerical examples of both linear and nonlinear systems validate the method. Lastly, in section 6 a sensitivity analysis is provided to investigate the full state measurement recommendation, while in section 7 the conclusions are drawn.

## 2. Formulation of the standard unscented Kalman filter

For the mathematical implementation of the standard UKF, consider the nonlinear process equation in the continuous time-domain and in a state space format:

$$\dot{\mathbf{z}}(t) = f\big(\mathbf{z}(t), \mathbf{u}(t)\big) + \boldsymbol{\nu}(t) \tag{1}$$

and the nonlinear observation equation:

$$\mathbf{y}(t) = h\big(\mathbf{z}(t), \mathbf{u}(t)\big) + \boldsymbol{\eta}(t) \tag{2}$$

The state vector $\mathbf{z}(t) = [\mathbf{x}(t), \dot{\mathbf{x}}(t), \boldsymbol{\theta}]^T$ includes the parameters of the system $\boldsymbol{\theta}$ apart from the dynamic states, while $\mathbf{y}(t)$ is the observation vector. Also, $f(\bullet)$ and $h(\bullet)$ are the state transition function and the observation function, respectively, which take into account the input vector $\mathbf{u}(t)$. Lastly, $\boldsymbol{\nu}(t)$ and $\boldsymbol{\eta}(t)$ are the process and the measurement noise, of covariance matrices $\mathbf{Q}(t)$ and $\mathbf{R}(t)$, respectively. Eqs. 1 and 2 can be discretized as:

$$\mathbf{z_k} = F(\mathbf{z_{k-1}}, \mathbf{u_{k-1}}) + \boldsymbol{\nu_{k-1}} \tag{3}$$

and,

$$\mathbf{y_k} = h(\mathbf{z_k}, \mathbf{u_k}) + \boldsymbol{\eta_k} \tag{4}$$

where,

$$F(\mathbf{z_{k-1}}, \mathbf{u_{k-1}}) = \mathbf{z_{k-1}} + \int_{(k-1)\Delta t}^{(k)\Delta t} f\big(\mathbf{z}(t), \mathbf{u}(t)\big)\ dt \tag{5}$$

Note that $k$ refers to $k\Delta t$ time instant, where $\Delta t$ is the sampling period. The discretized process and observation covariance matrices are:

$$\mathbf{Q_{k-1}} \approx \frac{\mathbf{Q}\big((k-1)\Delta t\big)}{\Delta t}, \quad \mathbf{R_k} = \frac{\mathbf{R}(k\Delta t)}{\Delta t} \tag{6}$$

Assuming additive Gaussian noise with zero mean, without loss of generality, the UKF steps are derived in Table 1A.

In this pseudo-code, $\lambda$ is given by $\lambda = \alpha^2(L+\kappa) - L$ with secondary parameter $\kappa = 0$ or $3-L$ [17], where $L$ is the dimension of $\mathbf{z_k}$. The constant $\alpha \in [10^{-4}, 1]$ determines the spread of the sigma points around $\mathbf{z_k}$, while the weights $V$ are given by:

$$\begin{aligned} V_0^m &= \frac{\lambda}{L+\lambda} \\ V_0^c &= \frac{\lambda}{L+\lambda} + (1-\alpha^2+\beta) \\ V_i^m &= V_i^c = \frac{\lambda}{2(L+\lambda)} \quad i = 1, ..., 2L \end{aligned} \tag{7}$$

where, $\beta$ is a constant that incorporates prior information of the $\mathbf{z_k}$ distribution.

## 3. An unknown input-parameter-state unscented Kalman filter

Simultaneously estimating the unknown input is possible using the UKF. Employing the predicted states at time step $k$, the input is estimated using the continuous equation of motion at the time instant $k\Delta t$ as:

$$\mathbf{u_k^e} \approx \mathbf{G}\left(\mathbf{\ddot{x}_k^m}, \mathbf{\dot{x}_k^p}, \mathbf{x_k^p}, ...\right) \tag{8}$$

where, $\mathbf{G}(\bullet)$ can be a linear or a nonlinear function. $\mathbf{G}(\bullet)$ also contains the estimated parameters and, thus, it is updated in every step.

Importantly, the predicted states are estimated using the equation $\mathbf{Z_p} = F(\mathbf{z_{k-1}}, \mathbf{u_{k-1}})$; with the prior input only. In the end, the known input rows of $\mathbf{u_k^e}$ are replaced with the known zero or non-zero inputs. This input is used in the updating process with the measurements $(\mathbf{Y_i} = \mathbf{H_k Z_{i,p}})$ in Table 1B.

However, this estimated $\mathbf{u_k^e}$ is erroneous since neither the predicted states have been updated with the measurements yet, nor the system parameters. This is generally acceptable and its noise characteristics are incorporated into the measurement noise of the observation equation. As a result, $\boldsymbol{\eta_k}$ models both the measurement noise and the input derived noise.

The final step is to correct further the input estimation using the updated with measurements dynamic states and parameters as:

$$\mathbf{u_k^e} \approx \mathbf{G}\big(\mathbf{\ddot{x}_k^m}, \mathbf{\dot{x}_k}, \mathbf{x_k}, ...\big) \tag{9}$$

and then, replace the known input rows of the final $\mathbf{u_k^e}$ estimation with the known zero or non-zero inputs.

The measurement error of acceleration still exists in contrast to the displacement, velocity, and parameter error but, since the process error is modeled, the success of the full estimation is not endangered.

This final $\mathbf{u_k^e}$ is successively used at the prediction calculation of the next step $k+1$, and the overall procedure is repeated.

Assuming additive Gaussian noise with zero mean, without loss of generality, the unknown input-parameter-state unscented Kalman filter (IPS-UKF) is derived in Table 1B, where $\lambda$, $\kappa$, $L$, $\alpha$,$\beta$ and $V$ have been defined in section 2.

Table 1: Standard and proposed UKF pseudo-codes.

| A. Joint parameter-state estimation using the UKF. | B. Joint input-parameter-state using the UKF (IPS-UKF). |
|---|---|
| step 1:<br>$k = 0$ (Time step)<br>$\mathbf{z_k} = E[\mathbf{z_0}]$ ($E$ stands for Expectation)<br>$\mathbf{P_k} = E[(\mathbf{z_0} - \mathbf{z_k})(\mathbf{z_0} - \mathbf{z_k})^T]$ (Covariance matrix) | step 1:<br>$k = 0$ (Time step)<br>$\mathbf{z_k} = E[\mathbf{z_0}]$<br>$\mathbf{P_k} = E[(\mathbf{z_0} - \mathbf{z_k})(\mathbf{z_0} - \mathbf{z_k})^T]$<br>$\mathbf{u_k^e} = E[\mathbf{u_0^e}]$ |
| step 2:<br>$\mathbf{Z_k} = [\mathbf{z_k} - \sqrt{(L+\lambda)\mathbf{P_k}}, \quad \mathbf{z_k}, \quad \mathbf{z_k} + \sqrt{(L+\lambda)\mathbf{P_k}}]$<br>$k = k + 1$<br>$\mathbf{Z_p} = F(\mathbf{Z_{k-1}}, \mathbf{u_{k-1}})$ (p stands for prediction)<br>$\mathbf{z_p} = \sum_{i=0}^{2L} V_i^m \mathbf{Z_{i,p}}$ | step 2:<br>$\mathbf{Z_k} = [\mathbf{z_k} - \sqrt{(L+\lambda)\mathbf{P_k}}, \quad \mathbf{z_k}, \quad \mathbf{z_k} + \sqrt{(L+\lambda)\mathbf{P_k}}]$<br>$k = k + 1$<br>$\mathbf{Z_p} = F(\mathbf{Z_{k-1}}, \mathbf{u_{k-1}^e})$<br>$\mathbf{z_p} = \sum_{i=0}^{2L} V_i^m \mathbf{Z_{i,p}}$<br>$\mathbf{u_k^e} \approx \mathbf{G}\big(\mathbf{\ddot{x}_k^m}, \mathbf{z_p}, ...\big)$ ($G(\bullet)$ involves estimated parameters and it is updated in every step)<br>Replace all $\mathbf{u_k^e}$ zero/non-zero known input rows |
| $\mathbf{P_p} = \sum_{i=0}^{2L} V_i^c [\mathbf{Z_{i,p}} - \mathbf{z_p}][\mathbf{Z_{i,p}} - \mathbf{z_p}]^T + \mathbf{Q_{k-1}}$<br>$\mathbf{Y_i} = h(\mathbf{Z_{i,p}}, \mathbf{u_k})$<br>$\mathbf{y} = \sum_{i=0}^{2L} V_i^m \mathbf{Y_i}$<br>$\mathbf{P_m} = \sum_{i=0}^{2L} V_i^c [\mathbf{Y_i} - \mathbf{y}][\mathbf{Y_i} - \mathbf{y}]^T + \mathbf{R_k}$<br>$\mathbf{P_s} = \sum_{i=0}^{2L} V_i^c [\mathbf{Z_{i,p}} - \mathbf{z_p}][\mathbf{Y_i} - \mathbf{y}]^T$ | $\mathbf{P_p} = \sum_{i=0}^{2L} V_i^c [\mathbf{Z_{i,p}} - \mathbf{z_p}][\mathbf{Z_{i,p}} - \mathbf{z_p}]^T + \mathbf{Q_{k-1}}$<br>$\mathbf{Y_i} = h(\mathbf{Z_{i,p}}, \mathbf{u_k^e})$<br>$\mathbf{y} = \sum_{i=0}^{2L} V_i^m \mathbf{Y_i}$<br>$\mathbf{P_m} = \sum_{i=0}^{2L} V_i^c [\mathbf{Y_i} - \mathbf{y}][\mathbf{Y_i} - \mathbf{y}]^T + \mathbf{R_k}$<br>$\mathbf{P_s} = \sum_{i=0}^{2L} V_i^c [\mathbf{Z_{i,p}} - \mathbf{z_p}][\mathbf{Y_i} - \mathbf{y}]^T$ |
| If $\mathbf{y_k^m}$ is the measurement vector at time step $k$:<br>$\mathbf{z_k} = \mathbf{z_p} + \mathbf{P_s}\mathbf{P_m}^{-1}(\mathbf{y_k^m} - \mathbf{y})$ (State estimation)<br>$\mathbf{P_k} = \mathbf{P_p} - \mathbf{P_s}\big(\mathbf{P_s}\mathbf{P_m}^{-1}\big)^T$ (Covariance estimation) | If $\mathbf{y_k^m}$ is the measurement vector at time step $k$:<br>$\mathbf{z_k} = \mathbf{z_p} + \mathbf{P_s}\mathbf{P_m}^{-1}(\mathbf{y_k^m} - \mathbf{y})$ (State estimation)<br>$\mathbf{P_k} = \mathbf{P_p} - \mathbf{P_s}\big(\mathbf{P_s}\mathbf{P_m}^{-1}\big)^T$ (Covariance estimation)<br>$\mathbf{u_k^e} \approx \mathbf{G}\big(\mathbf{\ddot{x}_k^m}, \mathbf{z_k}, ...\big)$ (Final Input estimation)<br>Replace all $\mathbf{u_k^e}$ zero/non-zero known input rows |
| step 3:<br>Go to step 2 until $k = k_{max}$ | step 3:<br>Go to step 2 until $k = k_{max}$ |

## 4. Identifiability discussion for the joint input-parameter-state estimation

At this point, a question arises about whether this system is identifiable. In the case where the input is known and there is no measurement noise in a linear system, the identifiability of the parameter vector can be checked using the local identifiability test [18]. For nonlinear systems, the local identifiability can be examined using the Observability Rank Condition [19].

In the case of the unknown input, however, the above tests do not hold. The SDOF linear system is not identifiable in the time-domain even if flawless knowledge of the mass and of all dynamic states is assumed.

A mathematical explanation of this phenomenon can be derived using the perturbation analysis. The sensitivity of an identification procedure is being studied given an erroneous damping parameter $c$, stiffness parameter $k$, or input $u(t)$. Instead of the true values: $c + \Delta c$, $k + \Delta k$ and $u(t) + \Delta u(t)$ are used, where $\Delta c$, $\Delta k$ and $\Delta u(t)$ denote the erroneous parts. By doing that, the linear SDOF equation of motion is modified as:

$$
\begin{aligned}
& m\ddot{x}(t) + (c + \Delta c)\dot{x}(t) + (k + \Delta k)x(t) = u(t) + \Delta u(t) : \\
& \bullet\, if \quad \Delta u(t) = 0 \implies -\frac{\Delta k}{\Delta c} = \frac{\dot{x}(t)}{x(t)} \quad \checkmark \\
& \bullet\, if \quad \Delta k = 0 \implies \Delta c = \frac{\Delta u(t)}{\dot{x}(t)} \quad \checkmark \\
& \bullet\, if \quad \Delta c = 0 \implies \Delta k = \frac{\Delta u(t)}{x(t)} \quad \checkmark
\end{aligned}
\tag{10}
$$

meaning that, even if the erroneous part of an unknown quantity is minimized, a second incorrectly identified combination exists that satisfies instantly the equality in the equation of motion.

Alternatively for the same conclusion, one could notice that:

$$\begin{aligned}
&m\ddot{x}(t) + c\dot{x}(t) + kx(t) + \Delta c\dot{x}(t) + \Delta kx(t) = u(t) + \Delta u(t) && \Longrightarrow \\
&m\ddot{x}(t) + c\dot{x}(t) + kx(t) = \Big\{u(t) + \Delta u(t) - \Delta c\dot{x}(t) - \Delta kx(t)\Big\} && \Longrightarrow \\
&m\ddot{x}(t) + c\dot{x}(t) + kx(t) = \overline{u}(t)
\end{aligned} \tag{11}$$

where, $\overline{u}(t) = u(t) + \Delta u(t) - \Delta c\dot{x}(t) - \Delta kx(t)$. By identifying different erroneous parts, an equivalent 'input' $\overline{u}(t)$ can always be found. As a consequence, systems with erroneous 'input' $\overline{u}(t)$, which have exactly the same mass and dynamic states, can be identified.

In the frequency domain, though, several methods provide a valid solution when no known (zero or non-zero) input is available, for example the frequency domain decomposition [20, 21]. Nonetheless, the input or the dynamic state estimation are not estimated, while also linear systems are only treated. The frequency domain decomposition especially is offline and it is difficult to automate.

However, a MDOF system with known zero or non-zero input at one DOF is potentially identifiable. Assume $\mathbf{C} + \Delta\mathbf{C}$, $\mathbf{K} + \Delta\mathbf{K}$, and $u_2(t) + \Delta u_2(t)$ for the perturbation analysis of a linear 2-DOF system, where $\Delta\mathbf{C}$, $\Delta\mathbf{K}$, and $\Delta u_2(t)$ denote the erroneous parts. By doing that, the linear 2-DOF equation of motion, assuming without loss of generality known zero input at DOF 1,

is modified as:

$$\mathbf{M}\begin{Bmatrix}\ddot{x_1}(t)\\ \ddot{x_2}(t)\end{Bmatrix}+\mathbf{C}\begin{Bmatrix}\dot{x_1}(t)\\ \dot{x_2}(t)\end{Bmatrix}+\mathbf{\Delta C}\begin{Bmatrix}\dot{x_1}(t)\\ \dot{x_2}(t)\end{Bmatrix}+\mathbf{K}\begin{Bmatrix}x_1(t)\\ x_2(t)\end{Bmatrix}+\mathbf{\Delta K}\begin{Bmatrix}x_1(t)\\ x_2(t)\end{Bmatrix}=\begin{Bmatrix}0\\ u_2(t)+\Delta u_2(t)\end{Bmatrix}\implies$$

$$\mathbf{M}\begin{Bmatrix}\ddot{x_1}(t)\\ \ddot{x_2}(t)\end{Bmatrix}+\mathbf{C}\begin{Bmatrix}\dot{x_1}(t)\\ \dot{x_2}(t)\end{Bmatrix}+\mathbf{K}\begin{Bmatrix}x_1(t)\\ x_2(t)\end{Bmatrix}=\left[\begin{Bmatrix}0\\ u_2(t)+\Delta u_2(t)\end{Bmatrix}-\mathbf{\Delta C}\begin{Bmatrix}\dot{x_1}(t)\\ \dot{x_2}(t)\end{Bmatrix}-\mathbf{\Delta K}\begin{Bmatrix}\dot{x_1}(t)\\ \dot{x_2}(t)\end{Bmatrix}\right]\implies \tag{12}$$

$$\mathbf{M}\begin{Bmatrix}\ddot{x_1}(t)\\ \ddot{x_2}(t)\end{Bmatrix}+\mathbf{C}\begin{Bmatrix}\dot{x_1}(t)\\ \dot{x_2}(t)\end{Bmatrix}+\mathbf{K}\begin{Bmatrix}x_1(t)\\ x_2(t)\end{Bmatrix}=\begin{Bmatrix}\overline{u}_1(t)\\ \overline{u}_2(t)\end{Bmatrix}$$

Here, one can identify another equation with equivalent erroneous 'input' $\overline{u}_2(t)$ that has exactly the same mass and dynamic states for the second row of Eq. 12. However, the equivalent $\overline{u}_1(t)$ 'input' would be wrong since its value is known and equal to zero. Therefore, this known input at the DOF 1, regardless of whether it is zero or non-zero, can potentially lead to correctly identify the real parameters and, as a result, the real input at DOF 2. The same holds for the nonlinear systems.

Interestingly, given an unrealistic calibration of the covariance parameters, the standard UKF is potential capable to tackle the case without any known input when this is incorporated into the process equation noise; i.e. white noise with mean value equal to 0.

## 5. Applications

### *5.1. Linear MDOF system*

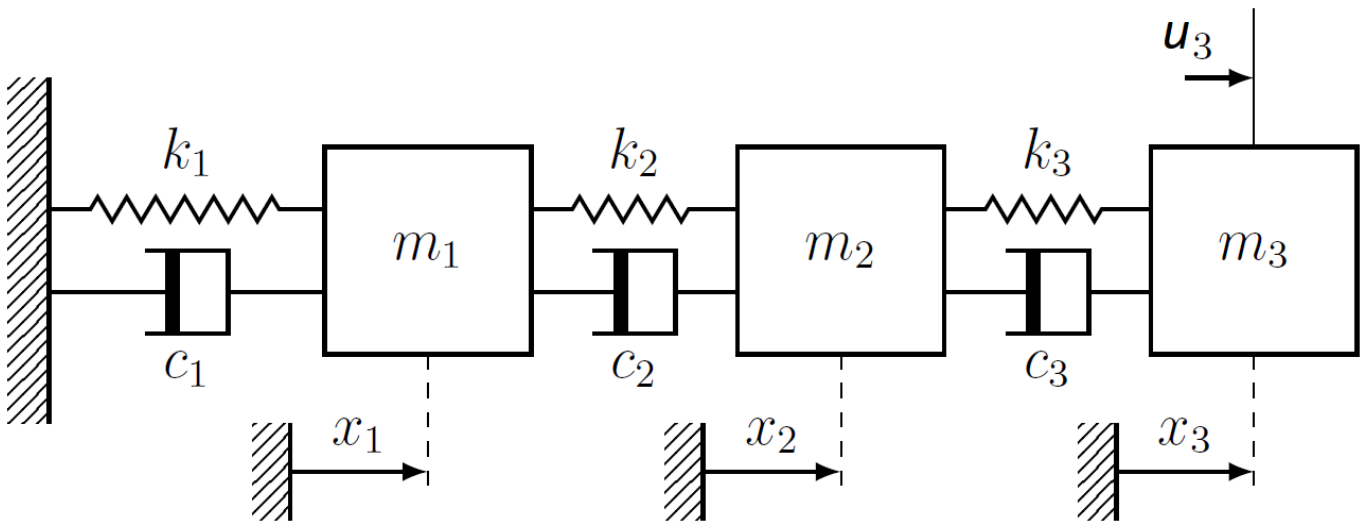

Figure 1: Linear 3-DOF sytem excited by a pulse at DOF 3.

For the numerical application of the proposed methodology consider the 3-DOF system of Fig. 1 which is described by the following equation:

$$\mathbf{M}\begin{Bmatrix}\ddot{x_1}(t)\\ \ddot{x_2}(t)\\ \ddot{x_3}(t)\end{Bmatrix}+\mathbf{C}\begin{Bmatrix}\dot{x_1}(t)\\ \dot{x_2}(t)\\ \dot{x_3}(t)\end{Bmatrix}+\mathbf{K}\begin{Bmatrix}x_1(t)\\ x_2(t)\\ x_3(t)\end{Bmatrix}=\begin{Bmatrix}0\\ 0\\ u_3(t)\end{Bmatrix} \tag{13}$$

where the system matrices which need to be identified (apart from $\mathbf{M}$) are:

$$\mathbf{M}=\begin{bmatrix}m_1 & 0 & 0\\ 0 & m_2 & 0\\ 0 & 0 & m_3\end{bmatrix}=\begin{bmatrix}1 & 0 & 0\\ 0 & 1 & 0\\ 0 & 0 & 1\end{bmatrix},\quad \mathbf{C}=\begin{bmatrix}c_1+c_2 & -c_2 & 0\\ -c_2 & c_2+c_3 & -c_3\\ 0 & -c_3 & c_3\end{bmatrix}=\begin{bmatrix}0.25+0.5 & -0.5 & 0\\ -0.5 & 0.5+0.75 & -0.75\\ 0 & -0.75 & 0.75\end{bmatrix},$$

$$\mathbf{K}=\begin{bmatrix}k_1+k_2 & -k_2 & 0\\ -k_2 & k_2+k_3 & -k_3\\ 0 & -k_3 & k_3\end{bmatrix}=\begin{bmatrix}9+11 & -11 & 0\\ -11 & 11+13 & -13\\ 0 & -13 & 13\end{bmatrix} \tag{14}$$

with initial conditions $\mathbf{x}(0)=[0\quad 0\quad 0]^T$ and $\dot{\mathbf{x}}(0)=[0\quad 0\quad 0]^T$. A pulse of $100N$ for $0.01s$ is applied at the time instant of $5s$, which is not known beforehand.

In order to create synthetic measurements, the Runge Kutta $4^{th}$ order method of integration is utilized to compute the system response for $30s$. The sampling frequency for the dynamic state measurements is considered to be 100 Hz, therefore the time discretization $\Delta t$ used in the Runge Kutta numerical solution is $0.01s$. Finally, to consider measurement noise, each response signal is contaminated by a Gaussian white noise sequence with a 5% root-mean-square noise-to-signal ratio.

In discrete time the system can be written, in an recursive form, as:

$$\mathbf{z_k} = \begin{bmatrix} \mathbf{x_{k-1}} + \Delta t \cdot \mathbf{\dot{x}_{k-1}} \\ \mathbf{\dot{x}_{k-1}} + \Delta t \cdot \mathbf{\ddot{x}_{k-1}} \\ \boldsymbol{\theta_{k-1}} \end{bmatrix} \tag{15}$$

and, using the equation of motion 13 to replace the accelerations $\mathbf{\ddot{x}_{k-1}}$, the process equation is written as:

$$\mathbf{z_k} = \begin{bmatrix} z_{1(k-1)} + \Delta t \cdot z_{4(k-1)} \\ z_{2(k-1)} + \Delta t \cdot z_{5(k-1)} \\ z_{3(k-1)} + \Delta t \cdot z_{6(k-1)} \\ z_{4(k-1)} + \Delta t \cdot m_1^{-1} \left\{ \begin{matrix} \cancelto{0}{u^e_{1(k-1)}} - \left(z_{7(k-1)} + z_{8(k-1)}\right) z_{4(k-1)} + z_{8(k-1)} z_{5(k-1)} \\ -\left(z_{10(k-1)} + z_{11(k-1)}\right) z_{1(k-1)} + z_{11(k-1)} z_{2(k-1)} \end{matrix} \right\} \\ z_{5(k-1)} + \Delta t \cdot m_2^{-1} \left\{ \begin{matrix} \cancelto{0}{u^e_{2(k-1)}} + z_{8(k-1)} z_{4(k-1)} - \left(z_{8(k-1)} + z_{9(k-1)}\right) z_{5(k-1)} + z_{9(k-1)} z_{6(k-1)} \\ + z_{11(k-1)} z_{1(k-1)} - \left(z_{11(k-1)} + z_{12(k-1)}\right) z_{2(k-1)} + z_{12(k-1)} z_{3(k-1)} \end{matrix} \right\} \\ z_{6(k-1)} + \Delta t \cdot m_3^{-1} \left\{ u^e_{3(k-1)} + z_{9(k-1)} z_{5(k-1)} - z_{9(k-1)} z_{6(k-1)} + z_{12(k-1)} z_{2(k-1)} - z_{12(k-1)} z_{3(k-1)} \right\} \\ z_{7(k-1)} \\ z_{8(k-1)} \\ z_{9(k-1)} \\ z_{10(k-1)} \\ z_{11(k-1)} \\ z_{12(k-1)} \end{bmatrix} \tag{16}$$

where,

$$\mathbf{z_k} = \begin{bmatrix} x_{1k} & x_{2k} & x_{3k} & \dot{x}_{1k} & \dot{x}_{2k} & \dot{x}_{3k} & c_{1k} & c_{2k} & c_{3k} & k_{1k} & k_{2k} & k_{3k} \end{bmatrix}^T \tag{17}$$

The presented formulation describes the Eq. 3 without the error vector $\boldsymbol{\nu_{k-1}}$,

while Eq. 4, without the error vector $\boldsymbol{\eta_k}$, can be written as:

$$\begin{bmatrix} x^m_{1(k)} \\ x^m_{2(k)} \\ x^m_{3(k)} \\ \dot{x}^m_{1(k)} \\ \dot{x}^m_{2(k)} \\ \dot{x}^m_{3(k)} \\ \ddot{x}^m_{1(k)} \\ \ddot{x}^m_{2(k)} \\ \ddot{x}^m_{3(k)} \end{bmatrix} = \begin{bmatrix} x^p_{1(k)} \\ x^p_{2(k)} \\ x^p_{3(k)} \\ \dot{x}^p_{1(k)} \\ \dot{x}^p_{2(k)} \\ \dot{x}^p_{3(k)} \\ \mathbf{M}^{-1}\left\{\mathbf{u^e_k} - \mathbf{C\dot{x}^p_k} - \mathbf{Kx^p_k}\right\} \end{bmatrix} \tag{18}$$

where for the unknown input $\mathbf{u^e_k} = \mathbf{G}\big(\mathbf{\ddot{x}^m_k}, \mathbf{\dot{x}^p_k}, \mathbf{x^p_k}, ...\big) = \mathbf{Mx^m_k} + \mathbf{C\dot{x}^p_k} + \mathbf{Kx^p_k}$ and then $\mathbf{u^e_k} \leftarrow [0 \quad 0 \quad u^e_{3(k)}]^T$.

The $\bullet^m$ and $\bullet^p$ stand for measured and predicted, respectively. The predicted states are available from Eq. 17. $\mathbf{C}$ and $\mathbf{K}$ matrices involve estimated parameters and have to be updated in every step.

Here, the full state measurement case is used. The investigation of this assumption is discussed in section 6.

The process covariance $\mathbf{Q_{k-1}}$ and the measurement covariance $\mathbf{R_k}$ matrices were chosen to be constant during the identification process and equal to $10^{-9} \cdot \mathbf{I_{12\times 12}}$ and $10^{-3} \cdot \mathbf{I_{9\times 9}}$, respectively. For larger values, the algorithm needs more data and time to converge, or it may even diverge.

In Fig. 2 the true and the estimated response, stiffness parameters, damping parameters, and input estimation and its error at DOF 3 are shown. The convergence is satisfactory.

Additionally, in Fig. 3 the application is shown for an ambient input;

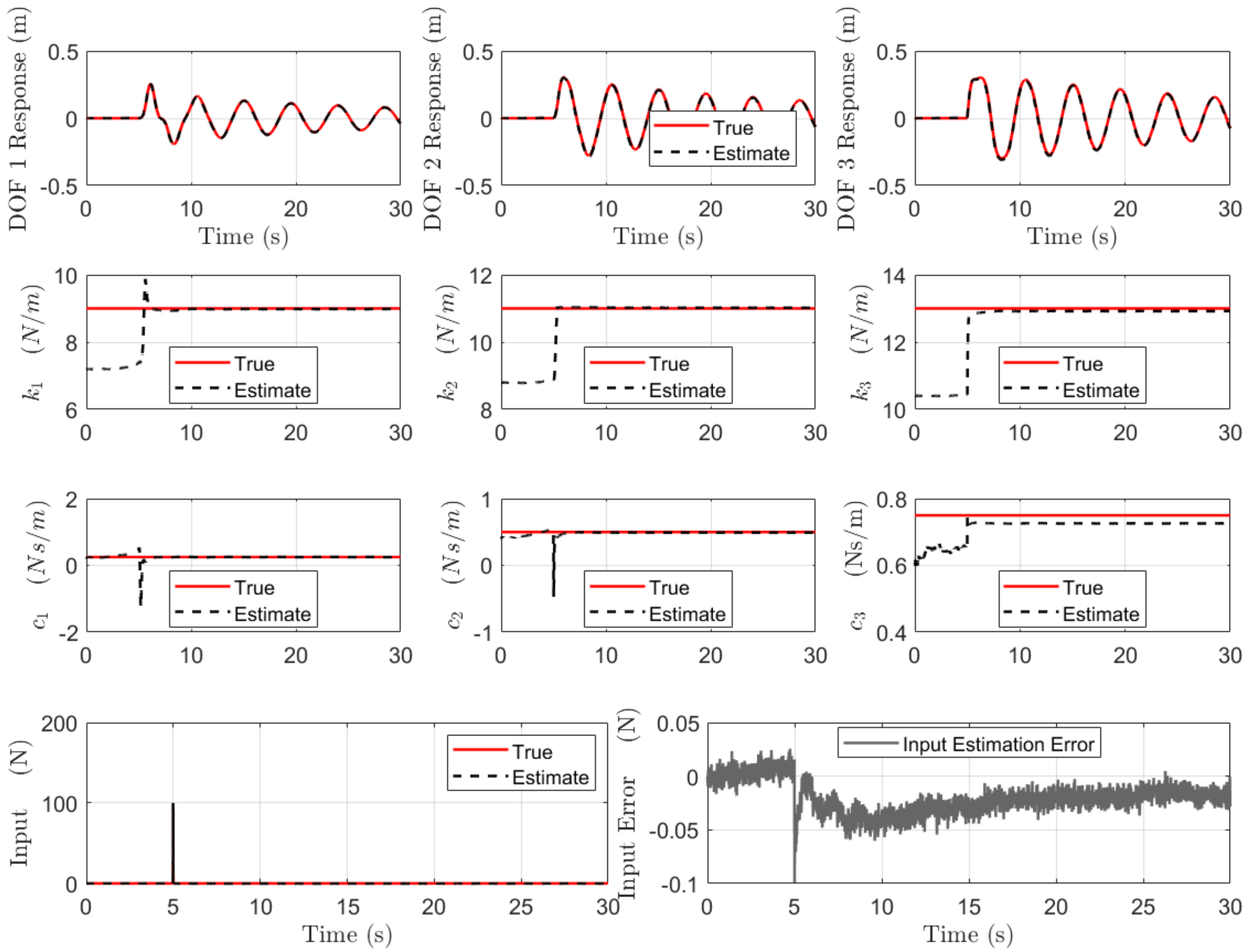

Figure 2: Results for the linear 3-DOF system. True and estimated response (first row), stiffness parameters (second row), damping parameters (third row), and input estimation and its error at DOF 3 (fourth row).

a noise-type input of mean value 0 and variance 4. The results are also satisfactory.

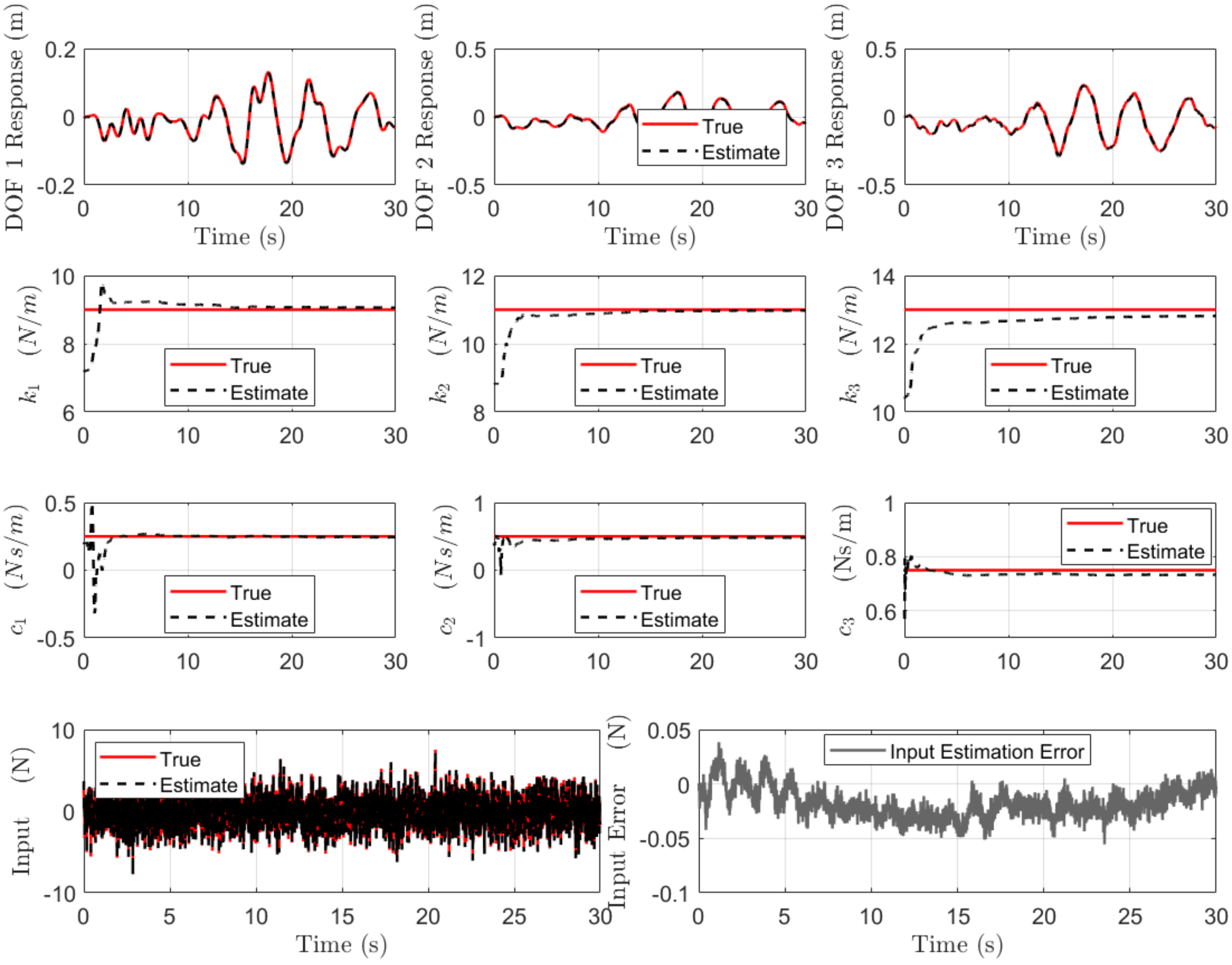

Figure 3: Results for the linear 3-DOF system under an ambient input. True and estimated response (first row), stiffness parameters (second row), damping parameters (third row), and input estimation and its error at DOF 3 (fourth row).

*5.2. Nonlinear MDOF system*

For a second numerical application, consider the Duffing nonlinear 2-DOF system described by the following equation:

$$\mathbf{M}\begin{Bmatrix}\ddot{x}_1(t)\\ \ddot{x}_2(t)\end{Bmatrix}+\mathbf{C}\begin{Bmatrix}\dot{x}_1(t)\\ \dot{x}_2(t)\end{Bmatrix}+\mathbf{K}\begin{Bmatrix}x_1(t)\\ x_2(t)\end{Bmatrix}+\mathbf{E}\begin{Bmatrix}x_1^3(t)\\ \left(x_2(t)-x_1(t)\right)^3\end{Bmatrix}=\begin{Bmatrix}0\\ u_2(t)\end{Bmatrix} \tag{19}$$

where the system matrices which need to be identified (apart from $\mathbf{M}$) are:

$$\mathbf{M}=\begin{bmatrix}m_1 & 0\\ 0 & m_2\end{bmatrix}=\begin{bmatrix}1 & 0\\ 0 & 1\end{bmatrix},\quad \mathbf{C}=\begin{bmatrix}c_1+c_2 & -c_2\\ -c_2 & c_2\end{bmatrix}=\begin{bmatrix}0.5+0.5 & -0.5\\ -0.5 & 0.5\end{bmatrix},$$

$$\mathbf{K}=\begin{bmatrix}k_1+k_2 & -k_2\\ -k_2 & k_2\end{bmatrix}=\begin{bmatrix}3+4.5 & -4.5\\ -4.5 & 4.5\end{bmatrix},\quad \mathbf{E}=\begin{bmatrix}\epsilon_1 & -\epsilon_2\\ 0 & \epsilon_2\end{bmatrix}=\begin{bmatrix}15 & -27\\ 0 & 27\end{bmatrix} \tag{20}$$

with initial conditions $\mathbf{x}(0)=[0\quad 0]^T$ and $\dot{\mathbf{x}}(0)=[0\quad 0]^T$. A pulse of $100N$ for $0.01s$ is applied at the time instant of $5s$, which is not known beforehand. Simultaneously, a noise-type input of mean value 0 and variance 4 is applied to the system. Both excitations are applied at DOF 2. The synthetic measurements are created in a similar manner to section 5.1.

In discrete time the system can be written, in an recursive form, as:

$$\mathbf{z_k}=\begin{bmatrix}\mathbf{x_{k-1}}+\Delta t\cdot\dot{\mathbf{x}}_{\mathbf{k-1}}\\ \dot{\mathbf{x}}_{\mathbf{k-1}}+\Delta t\cdot\ddot{\mathbf{x}}_{\mathbf{k-1}}\\ \boldsymbol{\theta_{k-1}}\end{bmatrix} \tag{21}$$

and, using the equation of motion 19 to replace the accelerations $\ddot{\mathbf{x}}_{\mathbf{k-1}}$, the process equation is written as:

$$
\mathbf{z_k} = \begin{bmatrix}
z_{1(k-1)} + \Delta t \cdot z_{3(k-1)} \\
z_{2(k-1)} + \Delta t \cdot z_{4(k-1)} \\
z_{3(k-1)} + \Delta t \cdot m_1^{-1} \left\{ \begin{array}{c} \cancelto{0}{u^e_{1(k-1)}} - \left(z_{5(k-1)} + z_{6(k-1)}\right) z_{3(k-1)} - \left(z_{7(k-1)} + z_{8(k-1)}\right) z_{1(k-1)} \\ + z_{6(k-1)} z_{4(k-1)} + z_{8(k-1)} z_{2(k-1)} - z_{9(k-1)} z^3_{1(k-1)} + z_{10(k-1)} \left(z_{2(k-1)} - z_{1(k-1)}\right)^3 \end{array} \right\} \\
z_{4(k-1)} + \Delta t \cdot m_2^{-1} \left\{ \begin{array}{c} u^e_{2(k-1)} + z_{6(k-1)} z_{3(k-1)} - z_{6(k-1)} z_{4(k-1)} + z_{8(k-1)} z_{1(k-1)} \\ - z_{8(k-1)} z_{2(k-1)} - z_{10(k-1)} \left(z_{2(k-1)} - z_{1(k-1)}\right)^3 \end{array} \right\} \\
z_{5(k-1)} \\
z_{6(k-1)} \\
z_{7(k-1)} \\
z_{8(k-1)} \\
z_{9(k-1)} \\
z_{10(k-1)}
\end{bmatrix} \tag{22}
$$

where,

$$
\mathbf{z_k} = \begin{bmatrix} x_{1k} & x_{2k} & \dot{x}_{1k} & \dot{x}_{2k} & c_{1k} & c_{2k} & k_{1k} & k_{2k} & \epsilon_{1k} & \epsilon_{2k} \end{bmatrix}^T \tag{23}
$$

The presented formulation describes the Eq. 3 without the error vector $\boldsymbol{\nu_{k-1}}$, while Eq. 4 without the error vector $\boldsymbol{\eta_k}$, can be written as:

$$
\begin{bmatrix} x^m_{1(k)} \\ x^m_{2(k)} \\ \dot{x}^m_{1(k)} \\ \dot{x}^m_{2(k)} \\ \ddot{x}^m_{1(k)} \\ \ddot{x}^m_{2(k)} \end{bmatrix} = \begin{bmatrix} x^p_{1(k)} \\ x^p_{2(k)} \\ \dot{x}^p_{1(k)} \\ \dot{x}^p_{2(k)} \\ \mathbf{M}^{-1} \left\{ \mathbf{u^e_k} - \mathbf{C}\dot{\mathbf{x}}^\mathbf{p}_\mathbf{k} - \mathbf{K}\mathbf{x}^\mathbf{p}_\mathbf{k} - \mathbf{E} \begin{bmatrix} \left(x^p_{1(k)}\right)^3 \\ \left(x^p_{2(k)} - x^p_{1(k)}\right)^3 \end{bmatrix} \right\} \end{bmatrix} \tag{24}
$$

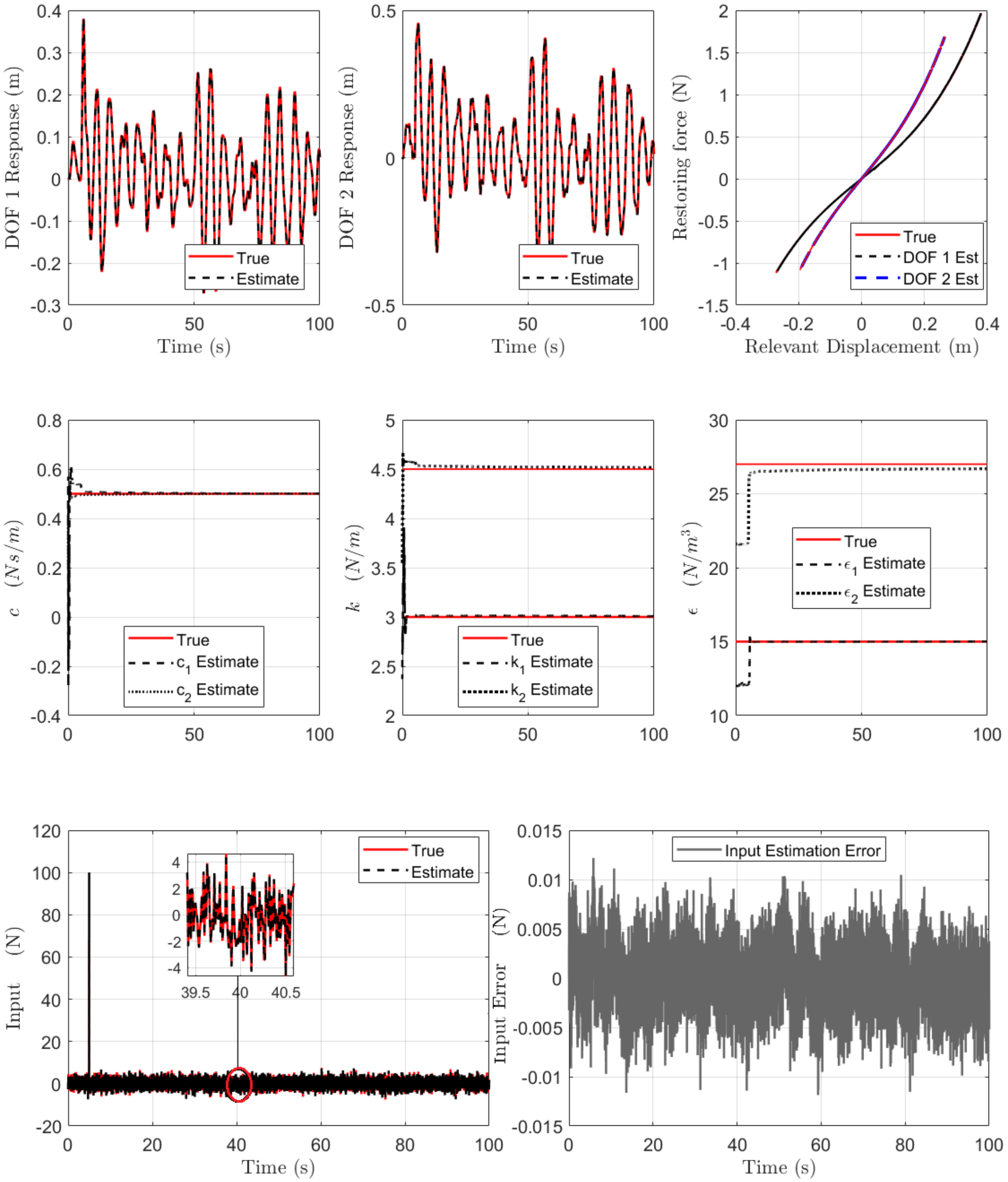

Figure 4: Results for the nonlinear 2-DOF system. True and estimated response (first row), parameters (second row), and input estimation and its error at the DOF 2 (third row). The nonlinear springs have been extensively excited (upper right plot).

where for the unknown input $\mathbf{u^e_k} = \mathbf{G}\big(\mathbf{\ddot{x}^m_k}, \mathbf{\dot{x}^p_k}, \mathbf{x^p_k}, ...\big) = \mathbf{M}\mathbf{x^m_k} + \mathbf{C}\mathbf{\dot{x}^p_k} + \mathbf{K}\mathbf{x^p_k} + \mathbf{E}\big[\big(x^p_{1(k)}\big)^3,\ \big(x^p_{2(k)} - x^p_{1(k)}\big)^3\big]^T$ and then $\mathbf{u^e_k} \leftarrow [0 \quad u^e_{2(k)}]^T$.

Here, the full state measurement is used. The investigation of this assumption, especially for this nonlinear system, is discussed in section 6.

The process covariance $\mathbf{Q_{k-1}}$ and the measurement covariance $\mathbf{R_k}$ matrices were chosen to be constant during the identification process and equal to $10^{-9} \cdot \mathbf{I_{10\times10}}$ and $10^{-5} \cdot \mathbf{I_{6\times6}}$, respectively.

In Fig. 4 the true and the estimated response, parameters, and input estimation and its error at DOF 2 are shown. The convergence is satisfactory.

## 6. Output information sensitivity analysis

The prior information recommendation is to measure all dynamic states, an assumption that sounds restrictive or economically suboptimal. A way to relax this requirement is via a data fusion algorithm (i.e. [22]), simultaneously with the IPS-UKF concept.

This is attributed to two main reasons: the high-frequency noise amplification when the non-measured derivative of a measured dynamic state is needed, and the low-frequency noise which causes integration error or the loss of the drift information when the non-measured integration of a measured dynamic state is needed.

The full state measurement recommendation is not mandatory though, as demonstrated using a sensitivity analysis. It is shown that measurements of at least two dynamic states at each DOF is potentially adequate to identify a nonlinear system.

Note that, the high-frequency noise is the hardest to tackle when the differentiation of a signal is needed, let alone when the input is unknown. For this reason, the investigation of the non-acceleration measurement case is abandoned for unknown input system identification.

Importantly, if the system is highly nonlinear, then the drift information loss will endanger the identification procedure when no displacements are measured.

Compared with the previous numerical application (section 5.2), only the observation equation is modified accordingly for the non-displacement, the non-velocity, and the acceleration-only measurement case, respectively, as:

$$\begin{bmatrix} \dot{x}^m_{1(k)} \\ \dot{x}^m_{2(k)} \\ \ddot{x}^m_{1(k)} \\ \ddot{x}^m_{2(k)} \end{bmatrix} = \begin{bmatrix} \dot{x}^p_{1(k)} \\ \dot{x}^p_{2(k)} \\ \mathbf{M}^{-1}\left\{ \mathbf{u^e_k} - \mathbf{C\dot{x}^p_k} - \mathbf{Kx^p_k} - \mathbf{E}\begin{bmatrix} \left(x^p_{1(k)}\right)^3 \\ \left(x^p_{2(k)} - x^p_{1(k)}\right)^3 \end{bmatrix} \right\} \end{bmatrix} \tag{25}$$

$$\begin{bmatrix} x^m_{1(k)} \\ x^m_{2(k)} \\ \ddot{x}^m_{1(k)} \\ \ddot{x}^m_{2(k)} \end{bmatrix} = \begin{bmatrix} x^p_{1(k)} \\ x^p_{2(k)} \\ \mathbf{M}^{-1}\left\{ \mathbf{u^e_k} - \mathbf{C\dot{x}^p_k} - \mathbf{Kx^p_k} - \mathbf{E}\begin{bmatrix} \left(x^p_{1(k)}\right)^3 \\ \left(x^p_{2(k)} - x^p_{1(k)}\right)^3 \end{bmatrix} \right\} \end{bmatrix} \tag{26}$$

and,

$$\begin{bmatrix} \ddot{x}^m_{1(k)} \\ \ddot{x}^m_{2(k)} \end{bmatrix} = \begin{bmatrix} \mathbf{M}^{-1}\left\{ \mathbf{u^e_k} - \mathbf{C\dot{x}^p_k} - \mathbf{Kx^p_k} - \mathbf{E}\begin{bmatrix} \left(x^p_{1(k)}\right)^3 \\ \left(x^p_{2(k)} - x^p_{1(k)}\right)^3 \end{bmatrix} \right\} \end{bmatrix} \tag{27}$$

Interestingly, the first two measurement cases can be characterized as satisfying. The response, all the parameters, and the input have been estimated adequately (Fig. 5 and Fig. 6).

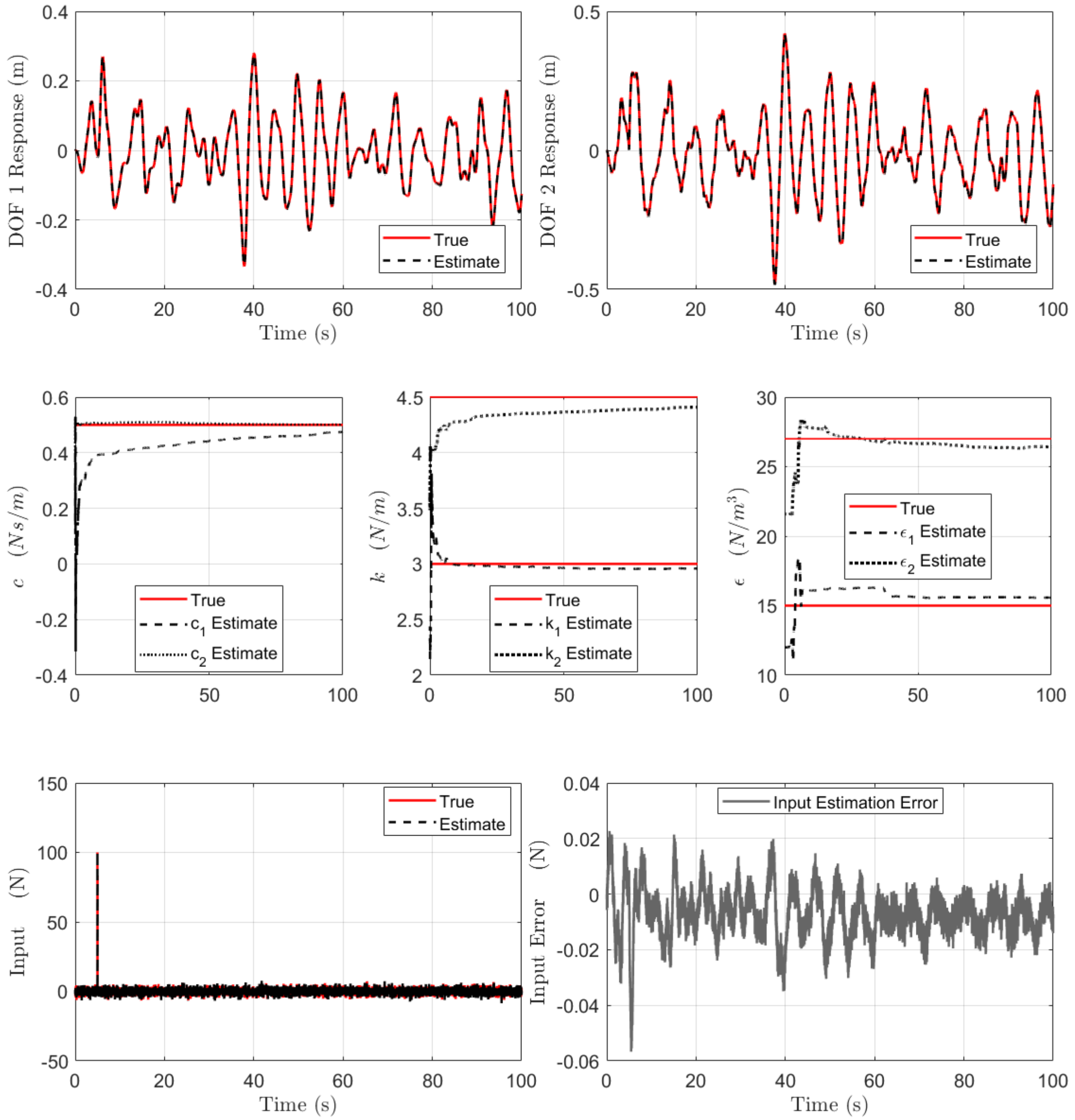

Figure 5: Results for the nonlinear 2-DOF system without displacements measurements. True and estimated response (first row), parameters (second row), and input estimation and its error at DOF 2 (third row).

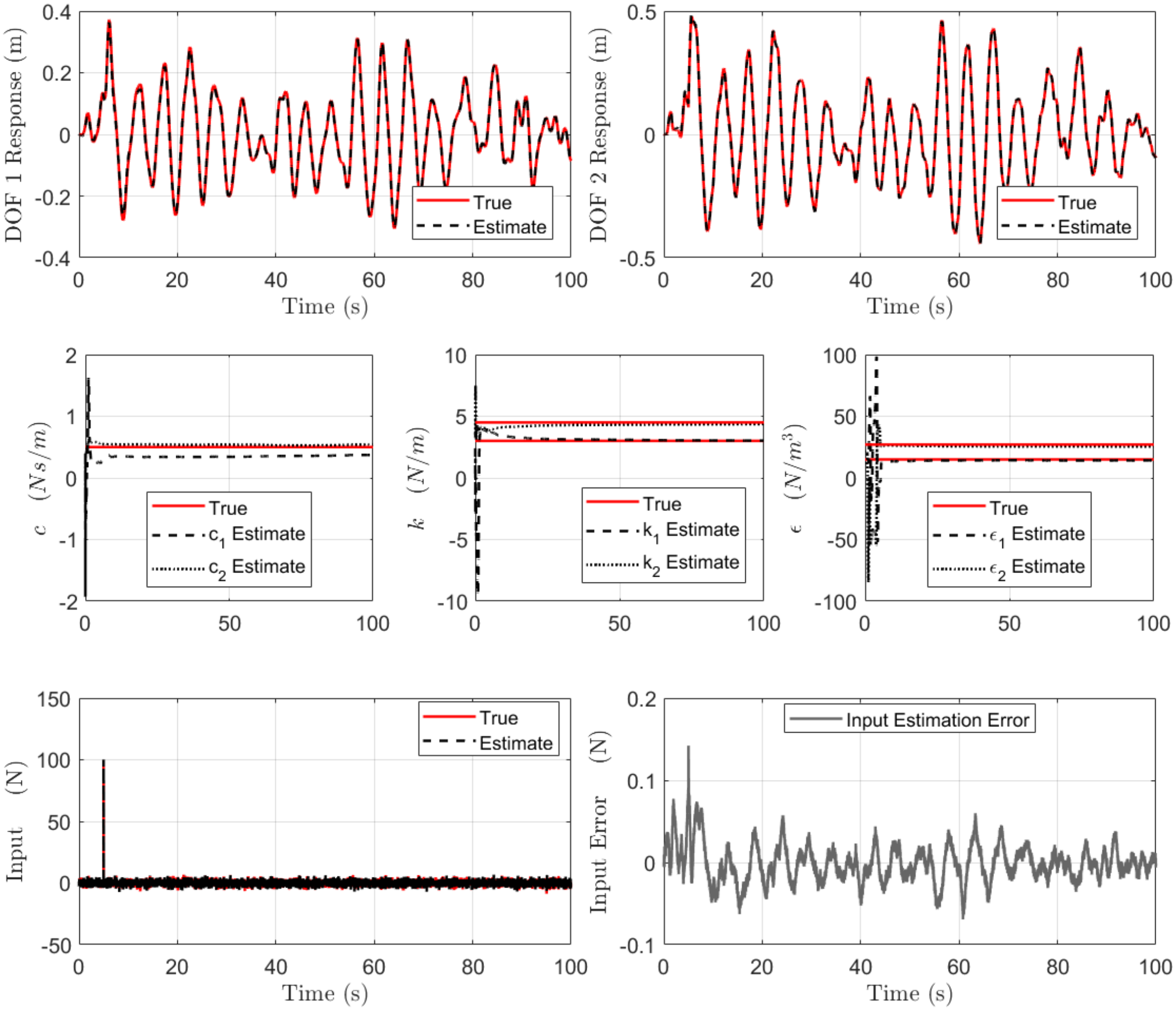

Figure 6: Results for the nonlinear 2-DOF system without velocity measurements. True and estimated response (first row), parameters (second row), and input estimation and its error at DOF 2 (third row)

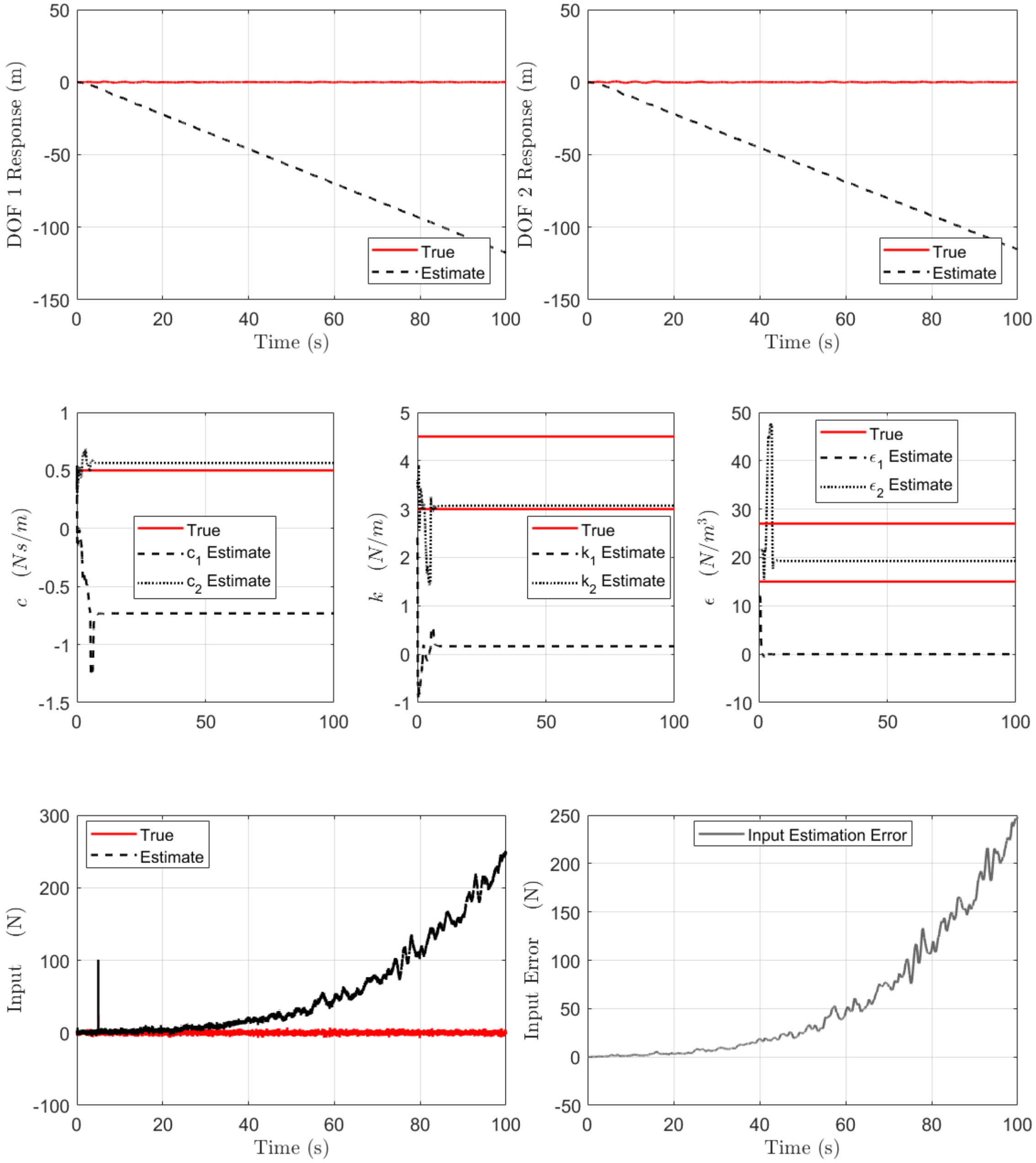

Figure 7: Results for the nonlinear 2-DOF system for the acceleration-only measurements case. True and estimated response (first row), parameters (second row), and input estimation and its error at DOF 2 (third row).

For the acceleration-only measurement case, the results were misleading and unreliable. Almost all the parameters and the dynamic states are not well estimated. Especially for the input and the response, a trend for continuously higher error is noticed (see Fig. 7). Noticeably, the unreliability of the acceleration-only measurement case is enhanced by the fact that for each IPS-UKF run, convergence to different values or even divergence happens.

The sensitivity investigation was done for the 2-DOF nonlinear system and thus, for a non-trivial identification problem since more DOFs potentially means more known zero inputs and, thus, faster and more accurate convergence.

## 7. Conclusions

The input-parameter-state estimation capabilities of a novel unscented Kalman filter for real time applications was examined herein, where the unknown input was estimated in two stages within each time step. A detailed Table was provided for the standard and the proposed unscented Kalman filter.

It was demonstrated using the perturbation analysis that a system with at least one zero or non-zero known input can potentially be uniquely identified.

Importantly, the process and measurement covariance matrices played an important role on the correct convergence of this methodology; the proper calibration is outside the scope of this work though.

For a successful full estimation displacement, velocity, and acceleration measurements are recommended, while a data fusion methodology may relax this assumption. However, this is not mandatory as was demonstrated via a sensitivity analysis.

Overall, this powerful procedure can successfully estimate the system response, inputs, and also identify the model parameters simultaneously and online.

**Acknowledgment**

The authors would like to acknowledge the support of the U.S. National Science Foundation, which partially supported this research under Grant No. CMMI-1563364.